\documentclass[12pt]{iopart}

\usepackage{iopams}
\usepackage{graphicx}% Include figure files
\usepackage{hyperref}% add hypertext capabilities
\begin{document}

\title{LAGRANGE: LAser GRavitational-wave ANtenna at GEo-lunar Lagrange points}

% author list

\author{J~W~Conklin$^1$,
  S~Buchman$^1$,
  V~Aguero$^6$,
  A~Alfauwaz$^3$,
  A~Aljadaan$^3$,
  M~Almajed$^3$,
  H~Altwaijry$^3$,
  T~Al-Saud$^3$,
  K~Balakrishnan$^1$,
  R~L~Byer$^1$,
  K~Bower$^5$,
  B~Costello$^5$,
  G~D~Cutler$^1$,
  D~B~DeBra$^1$,
  D~M~Faied$^2$,
  C~Foster$^2$,
  A~L~Genova$^2$,
  J~Hanson$^4$,
  K~Hooper$^5$,
  E~Hultgren$^1$,
  B~Jaroux$^2$,
  A~Klavins$^5$,
  B~Lantz$^1$,
  J~A~Lipa$^1$,
  A~Palmer$^5$,
  B~Plante$^5$,
  H~S~Sanchez$^2$,
  S~Saraf$^1$,
  D~Schaechter$^5$,
  T~Sherrill$^5$,
  K-L~Shu$^5$,
  E~Smith$^5$,
  D~Tenerelli$^5$,
  R~Vanbezooijen$^5$,
  G~Vasudevan$^5$,
  S~D~Williams$^6$,
  S~P~Worden$^2$,
  J~Zhou$^1$ and
  A~Zoellner$^1$
  }
\address{$^1$ W.W. Hansen Experimental Physics Lab,
  Stanford University, Stanford, CA 94305}

\address{$^2$ NASA Ames Research Center, Moffett Field, CA 94035}

\address{$^3$ King Abdulaziz City for Science and Technology, Riyadh,
  Saudi Arabia 11442}

\address{$^4$ CrossTrac Engineering, Inc. Sunnyvale, CA 94089}

\address{$^5$ Lockheed Martin Space Systems Company, Palo Alto, CA 94304}

\address{$^6$ SRI International, Menlo Park, CA 94025}

\ead{johnwc@stanford.edu}

\begin{abstract}
We describe a new space gravitational wave observatory design called LAGRANGE
that maintains all important LISA science at about half the cost and with
reduced technical risk.
It consists of three drag-free spacecraft in the most stable geocentric
formation, the Earth-Moon L3, L4, and L5 Lagrange points.
Fixed antennas allow continuous contact with the Earth,
solving the problem of communications bandwidth and latency.
A 70 mm diameter AuPt sphere with a 35 mm gap to its enclosure
serves as a single inertial reference per 
spacecraft, operating in ``true'' drag-free mode
(no test mass forcing).
This is the core of the Modular Gravitational Reference Sensor
whose other advantages are:
a simple caging design based on the DISCOS 1972 drag-free mission,
an all optical read-out with pm fine and nm coarse sensors, and the extensive
technology heritage from the Honeywell gyroscopes, and the
DISCOS and Gravity Probe B drag-free sensors.
An Interferometric Measurement System,
designed with reflective optics and a highly stabilized frequency standard,
performs the ranging between test masses and
requires a single optical bench with one laser per spacecraft.
Two 20 cm diameter telescopes per spacecraft, each with in-field pointing,
incorporate
novel technology developed for advanced optical systems by Lockheed Martin,
who also designed the spacecraft based on a multi-flight
proven bus structure.
Additional technological advancements include the drag-free propulsion,
thermal control, charge management systems, and materials.
LAGRANGE sub-systems are designed to be scalable and modular,
making them interchangeable with those of LISA or other gravitational
science missions.
We plan to space qualify critical technologies on small and nano
satellite flights, with the first launch (UV-LED Sat) in 2013.

\end{abstract}

%\maketitle
\maketitle

\section{Introduction}

Stanford University, NASA Ames Research Center, Lockheed Martin,
the King Abdulaziz City for Science and Technology (KACST), and
SRI International have formed a collaboration
(called SALKS)
to develop a new space gravitational-wave observatory mission concept,
named LAGRANGE, which
maintains all important LISA science at reduced cost
and with reduced technical risk.
We achieve this goal by revisiting all aspects of LISA for possible
improvements, while structuring the new elements to be modular and
scalable, as well as interchangeable with baseline LISA systems.
We incorporate both new technologies developed after the LISA and
LISA pathfinder designs were baselined
(UV-LEDs, non-transmissive optics, SRI thrusters,
test mass (TM) coatings and others),
as well as older space qualified technologies from
Honeywell (1953), DISCOS (1972) \cite{triad1974},
GP-B (2004) \cite{everitt2011}. 

LAGRANGE comes close to meeting the LISA sensitivity below
10 mHz and exceeds it at higher frequencies (see Fig. \ref{fig:strain}).
An internal NASA cost analysis,
cross checked against previous mission data,
gives a Rough-Order-of-Magnitude (ROM) cost of \$950M, with 30\% margin,
significantly less than the current LISA cost \cite{astro2010},
while including many technical advantages.

The three main elements of a space-based gravitational-wave observatory are
1) the constellation and its orbit 2) the gravitational reference sensors,
and 3) the metrology system
(see Table \ref{tab:compare} for comparison of LAGRANGE to LISA).
The following are the discriminating improvements of LAGRANGE.

1. LAGRANGE consists of a triangular constellation of identical spacecraft
(S/C) and payload at the Earth-Moon (E-M) L3, L4, and L5 Lagrange points.
This is the most stable geocentric configuration.
% with an arm length greater than 0.3~Gm.
Launch of the three spacecraft with one small propulsion module is possible
on a Falcon 9 at a cost of \$118M from NASA Launch Services.
Earth-based receivers are continuously in the field of view of fixed
transmitters on each spacecraft increasing the communication bandwidth by
$>$100 from LISA and greatly reducing
data latency (minutes instead of days).
From the experience with GP-B and LIGO,
the closest analogs to date,
this large bandwidth is an absolute requirement for mission success.

2. The single Gravitational Reference Sensor (GRS) is based on the Modular
Gravitational Reference Sensor (MGRS) concept
developed by SALKS and consists of a spherical test mass spinning at
3-10 Hz thus providing frequency separation from the 1 mHz to 1 Hz
primary LAGRANGE bandwidth.
Similar TMs have been successfully flown on DISCOS, and GP-B.
The MGRS is ‘true’ drag-free with no forcing in any direction,
and has a 35 mm gap that can be increased if necessary.
Caging by a single screw mechanism was demonstrated on the DISCOS
flight and is critical to the risk reduction in LAGRANGE versus LISA.
Magnetic spin-up and polhode damping were demonstrated thoroughly in
Honeywell gyroscopes. 

3. The Interferometric Measurement System (IMS)
consists of a single laser and optics bench that
incorporates only reflecting elements in the critical locations:
gratings and mirrors, while laser frequency stabilization is enhanced
by high finesse optical cavities and/or iodine molecular clocks.
Two telescopes per S/C with in-field pointing are designed to
minimize path-length errors.
% The six telescopes can be activated sequentially helping to separate
% pointing errors from the signal.

\begin{table}%[H] add [H] placement to break table across pages
  \caption{\label{tab:compare} Top-level LISA / LAGRANGE comparison.}
  \begin{tabular}{l c c}
    \hline
    & LISA & LAGRANGE \\
    \hline
    Number of spacecraft & 3 & 3 \\
    Orbit & heliocentric, 20$^\circ$ Earth trailing & Earth-Moon L3, L4, L5 \\
    Wet launch mass & $\sim$5,000 kg & 2,070 kg \\
    Arm length & 5 Gm & 0.67 Gm \\
    IMS sensitivity & $\mathrm{18 \ pm \:Hz^{-1/2}}$
      & $\mathrm{5 \ pm \:Hz^{-1/2}}$ \\
    DRS accel. noise & $\mathrm{3 \ fm/s^2 \:Hz^{-1/2}}$
      & $\mathrm{3 \ fm/s^2 \:Hz^{-1/2}}$ \\
    Observation period & 5 yr & 5 yr \\
    Telescopes / spacecraft & 2 $\times$ 40 cm & 2 $\times$ 20 cm \\
    GRSs / spacecraft & 2 & 1 \\
    Optics benches/spacecraft & 2 & 1 \\
    Laser power/spacecraft & 2 $\times$ 1.2 W & 1 $\times$ 1 W \\
    Controlled degrees of & 19 & 7 \\
    freedom / spacecraft & & \\
    Beam steering & articulated & in-field \\
    & optics \& GRS & pointing \\
    \hline
  \end{tabular}
\end{table}

Significant cost reduction for LAGRANGE over LISA was
achieved principally in two ways:
(a) by decreasing the per spacecraft mass and power by
reducing payload components
(2 lasers, 2 GRSs, and 2 optics benches for LISA
was reduced to 1 of each for LAGRANGE),
and (b) by using a geocentric orbit, which
requires only one propulsion module for all three spacecraft
and reduces mission operations complexity by increasing communications
bandwidth.

% Some challenges have been identified,
% but these appear to be solvable, building on the tremendous
% amount of development work already performed.
% It maintains all the important LISA science goals with an estimated cost of
% $\sim$ \$700M and could be pursued on a much faster launch schedule than
% that currently slated for LISA.
There is incremental risk reduction in LAGRANGE
due to an emphasis on simplicity.
In addition, the SALKS collaboration is implementing flight demonstrations of
critical technologies on small satellites and CubeSats \cite{worden2010}.
These include charge management, laser frequency stabilization,
shadow and interferometric position measurement, thrusters
and caging mechanisms.
The charge management flight (UV LED Sat \cite{sun2011}) is scheduled for 2013.
% The discussion below is focused mainly on justifying these claims,
% taking as a given that the science case has already been adequately
% demonstrated.

The structure of this paper is as follows: we start with a very brief
summary of the science capabilities;
follow with the design overview and science orbit;
discuss LAGRANGE specific details of the proposed instrument,
focusing on the IMS
and the Disturbance Reduction System (DRS);
describe the mission and spacecraft design, and the cost estimate;
closing with a discussion of TRLs.

\section{Science Capabilities}

% [0.5 pages - Stanford (Sasha)]
%
\vspace{-10pt}
The primary measurement band for LAGRANGE is 1~mHz to 1~Hz,
where the strain sensitivity is $3 \times 10^{-20}$.
The target astrophysical sources include:
% (Sasha, these are taken from the RFI. I simply reduced the frequency ranges):

\begin{enumerate}

  \item Massive black hole mergers in the range of $10^4$ (MBH) to $10^7$
    (SMBH) solar
    masses with orbit periods of $10^2$ to $10^4$ sec, giving
    signal-to-noise ratios (SNR) up to several thousands out to $z \sim 15$.

  \item Merging of stellar mass compact objects with massive black holes
    (EMRI) with signal periods of $10^2$ to $10^3$ seconds.

  \item Stellar mass binaries within the Milky Way with orbital periods
    of $10^2$ to $10^3$ seconds.

\end{enumerate}

\begin{figure}
  \begin{center}
  \includegraphics[width = 10 cm]{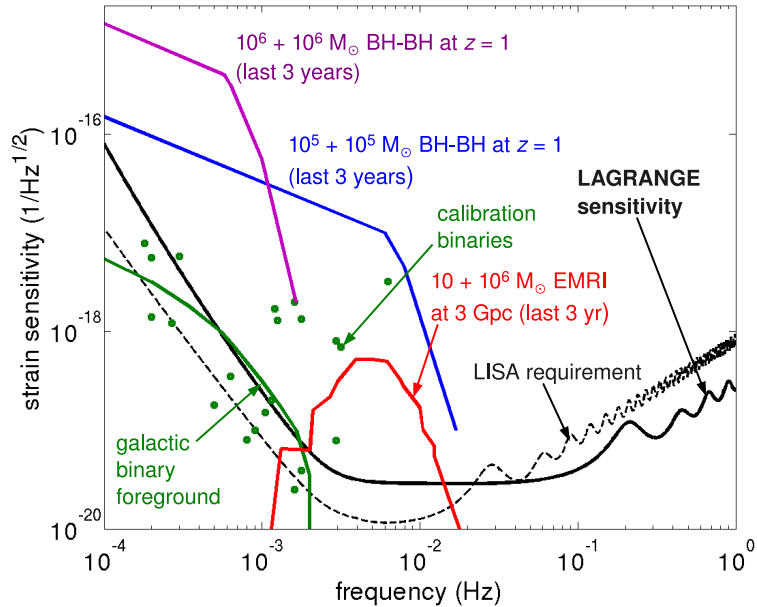}
  \caption{LAGRANGE strain sensitivity, and comparison with LISA
    requirements and gravitational wave sources,
    adapted from \cite{stebbins2009}.
    \label{fig:strain}}
  \end{center}
\end{figure}

Figure \ref{fig:strain} shows the estimated LAGRANGE strain
sensitivity (solid black curve) in units of $\mathrm{Hz^{-1/2}}$,
compared to the LISA requirement (dashed curve).
Colored curves and points represent the various known sources within the
LAGRANGE bandwidth.
The green curve is the confusion noise from unresolved galactic binaries
that dominates instrumental noise between $5\times10^{–4}$ and
$2\times10^{–3}$ Hz.
All sources above the sensitivity curve are detectable by LAGRANGE.
The green points represent the frequencies and strengths of known
Galactic binaries; their height above the noise curve
gives their SNR.
The purple, blue and red curves represent sources
(two SMBH binaries, and an EMRI, respectively)
whose frequency evolves upward during LAGRANGE's observation.

Sensitivity normal to the ecliptic plane is less than that of LISA
due to the reduced out-of-plane motion of the observatory.
However, higher gravitational-wave harmonics provide a significant
improvement in the position determination of MBH binaries \cite{wainright2009}.
Locating spinning black holes in a black hole binary is
much more accurate than would be expected from the
modulation produced by LISA’s precessing plane
alone \cite{lang2009, wainright2009}.

LAGRANGE will achieve the three most important science
goals of LISA listed in the 2010 astrophysics decadal survey,
``New Worlds, New Horizons'' \cite{astro2010}:

\begin{enumerate}
  \item Measurements of black hole mass and spin will be important for
    understanding the significance of mergers in the building of galaxies;
  \item Detection of signals from stellar-mass compact stellar remnants as
    they orbit and fall into massive black holes would provide exquisitely
    precise tests of Einstein's theory of gravity; and
  \item Potential for discovery of waves from unanticipated or exotic sources,
    such as backgrounds produced during the earliest moments of the
    universe or cusps associated with cosmic strings.
\end{enumerate}

It is therefore clear that if the technology goals are met,
LAGRANGE will deliver excellent science, little reduced from LISA.

\section{Observatory Design Overview}

\begin{figure}
  \begin{center}
  \includegraphics[width = 10 cm]{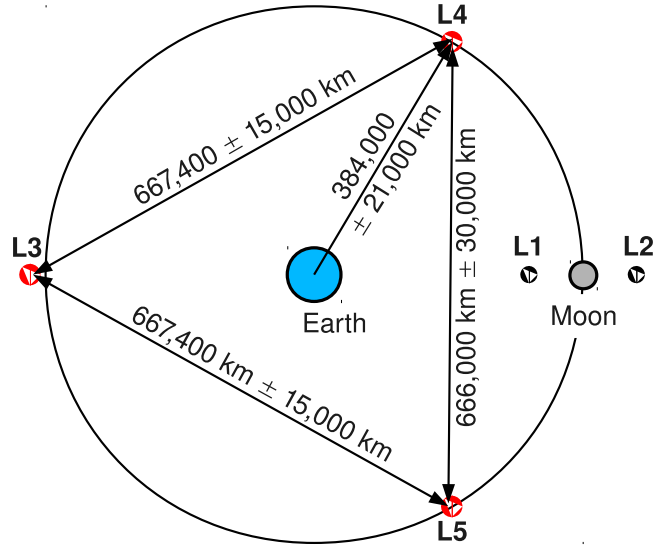}
  \caption{Orbit design with 3 drag-free spacecraft at the Earth-Moon L3, L4,
    and L5 LAGRANGE points.}
    \label{fig:orbit}
  \end{center}
\end{figure}

% [0.5 pages - Stanford (Sasha, Conklin)]
%
\vspace{-10pt}
LAGRANGE consists of a triangular constellation of three identical
spacecraft at the L3, L4, and L5 Lagrange points of the Earth-Moon
system as shown in Fig.~\ref{fig:orbit}.
This is the most stable geocentric configuration and has an average arm
length of 670,000 km.
Detection and observation of gravitational waves is performed using
laser interferometry to measure the distances between inertial references
in each spacecraft as in LISA \cite{benderLISAprePhaseA1998}.
Each spacecraft contains a single spherical test mass
as the inertial reference and
a single optical bench serving as a metrology reference.
The light source is a 1 W 1064 nm wavelength laser,
while two 20 cm aperture telescopes send and
receive laser light to and from the remote spacecraft.

The fundamental measurement scheme is based on the LISA approach.
The interferometric science measurement is made in two steps.
The first is the short-arm interferometer, which measures the optical bench
position with respect to the TM center of mass. 
The second is the long-arm interferometer that measures the distance
from the local optical bench to the optical bench on the remote spacecraft.
Time delay interferometry \cite{armstrong1999}
combines phase measurements made on-board each spacecraft,
accounting for the light travel time between spacecraft to 
cancel laser noise while retaining the gravitational wave signal.

% The following sections provide details of the science orbit,
% Interferometric Measurement System (IMS),
% and the Disturbance Reduction Systems (DRS).

\section{Science Orbit}

% [0.5 page - NASA ARC (write-up by Conklin with input from
%   Hugo, Anthony, Cyrus)]
%
\vspace{-10pt}
While the Earth-Moon Lagrange points provide the most stable geocentric
orbits, the gravitational attraction of the sun generates some instability.
The initial position and velocity of each spacecraft
have been chosen to maximize the time between station keeping maneuvers
of $<$~1~m/s once every 6-12 months,
performed using the on-board micronewton propulsion system.

Table \ref{tab:orbit} compares the stability and orbital dynamics
of the Earth-Moon L3, L4, L5 orbit with that of LISA.
Further improvement in the stability of the LAGRANGE constellation is expected
through simultaneous optimization of the initial conditions
for all three spacecraft,
minimizing range rate and breathing angle variations.
The spacecraft at L3 follows a perturbed halo orbit, roughly 50,000 km
in diameter and canted with respect to the plane of the Moon's orbit by
$\sim45^\circ$.
The spacecraft at L4 and L5 follow semi-periodic orbits as well,
but with more complex geometries.
One of the alternatives under study is to place all three spacecraft
in periodic orbits with similar phases, thus reducing
range rate variations.
% It also may be possible to generate more out-of-plane motion to
% improve gravitational wave observability in all directions.

\begin{figure}
  \begin{center}
  \includegraphics[width = 10 cm]{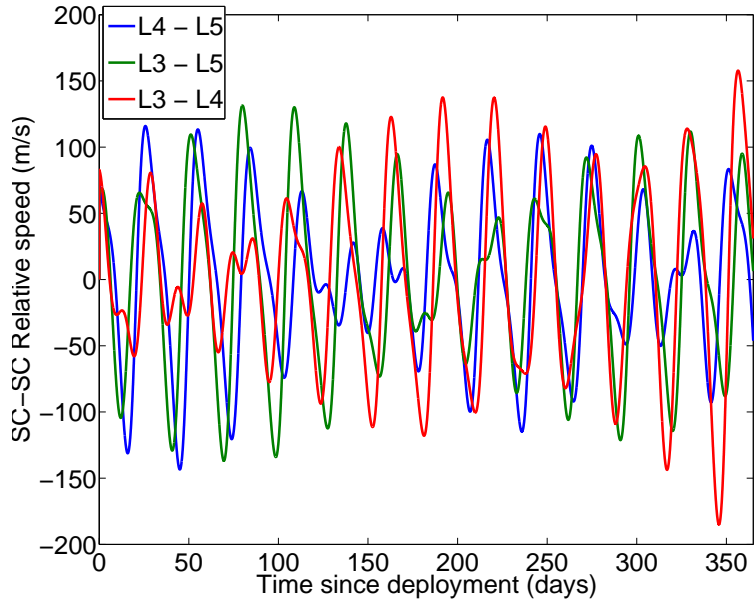}
  \caption{LAGRANGE S/C-to-S/C range rates over 1 year.
    \label{fig:L345doppler}}
  \end{center}
\end{figure}

The initial orbit design exhibits dynamics 
5 to 10 times larger than LISA.
Range rates between spacecraft vary by
$\pm$150 m/s, as shown in Fig. \ref{fig:L345doppler}.
This means that if the transmitted laser frequency is held constant,
the on-board phase measurements must accommodate heterodyne
frequencies of $\lesssim$150 MHz, compared
with $<$ 20 MHz for LISA.
However, since the range rates to the two remote spacecraft exhibit
large common mode variations (see Fig. \ref{fig:L345doppler}),
tuning the laser frequency on each spacecraft to the mean of the two known
Doppler frequencies, reduces the heterodyne frequency to $\lesssim$ 50 MHz.
The telescope must accommodate $\pm$ 2.5 deg of beam steering for the IMS
to remain locked to the remote spacecraft.

\begin{table}%[H] add [H] placement to break table across pages
  \caption{\label{tab:orbit} Comparison of E-M L3, L4, L5 and LISA orbits.}
  \begin{tabular}{l c c}
    \hline
    Parameter & E-M L3, L4, L5 & LISA \\
    \hline
    Nominal arm length & 670,000 km & 5,000,000 km \\
    Max. arm length variation & $\lesssim$ 5 \% & 1\% \\
    Breathing angle range & $\pm\lesssim$ 5 deg & $\pm$0.5 deg \\
    Max. S/C-to-S/C range rate & $\lesssim$ 150 m/s & 10 m/s \\
    Variation of orbit plane & 5 deg & 60 deg \\
    \hline
    % Launch C3 & $-1.7$ $\mathrm{kg^2/s^2}$ & 0.3 $\mathrm{kg^2/s^2}$ \\
  \end{tabular}
\end{table}
%
% \begin{figure}
%   \begin{center}
%  %\includegraphics[width = 8 cm]{}
%  \caption{(a) Spacecraft range variations, (b) Range rate variations,
%    (c) spacecraft angle variations
%    \label{fig:orbitVar}}
%  \end{center}
% \end{figure}

\section{Interferometric Measurement System}

\subsection{Interferometry}

% [0.5 pages - Stanford (Shally, Lipa, Byer, Lantz, Hollberg, Conklin)]
%
The IMS has two main components: The short-arm interferometer
determines the distance from the optics bench to the mass center
of the TM, and the long-arm interferometer measures the distance
between the optics benches on two spacecraft.
The combined TM-to-TM one-way measurement accuracy is estimated to be
$\mathrm{8 \ pm \: Hz^{-1/2}}$
($\mathrm{4 \ pm \: Hz^{-1/2}}$ shot noise limit).
While the internal interferometer measures distances that vary by
less than one wavelength (1 $\mu$m) during science operations, the external
interferometer must track laser phase over $\sim10^{14}$ wavelengths and
accommodate Doppler shifts of $\lesssim$ 150 MHz due to
S/C-to-S/C range and range rate variations.

Both long and short-arm interferometers are supplied by a
single 1 W Nd:YAG Nonplanar Ring Oscillator (NPRO) laser that is fiber
linked to the optical bench.
Once on the bench, all metrology is done with free-space optics.
Two 20 cm aperture telescopes
per spacecraft send and receive laser light to and from
the remote spacecraft.
A phasemeter is used to measure the long-arm interferometer phase
to 1 $\mathrm{\mu cycle \: Hz^{-1/2}}$ in the measurement band.

% \subsection{Optical Bench}

\begin{figure}
  \begin{center}
  \includegraphics[width = 16 cm]{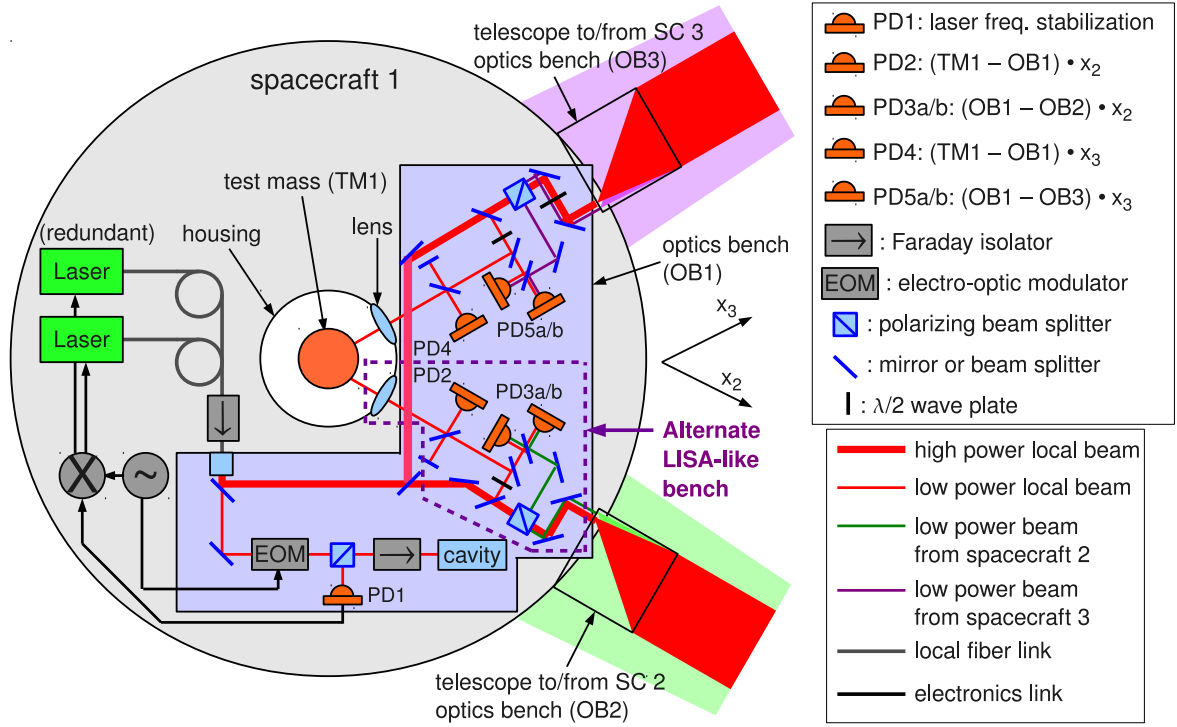}
  \caption{IMS and MGRS block diagram, with the alternate LISA-like
    optics bench. The primary grating based design replaces all components
    inside the dotted lines.
    \label{fig:IMSlayout}}
  \end{center}
\end{figure}

% [0.5 page - Stanford (Shally, Lipa, Byer, Lantz, Hollberg, Conklin)]
%
\emph{Optical Bench}:
Once on the optics bench, a small portion of the 1 W beam is separated off
and fed into the frequency stabilization system, shown as the
Electro-Optic Modulator (EOM) plus
optical cavity in Fig. \ref{fig:IMSlayout}.
The main beam then passes through a 50/50 beam splitter, which
divides the power between the right and left arms.

\begin{figure}
  \begin{center}
  \includegraphics[width = 11 cm]{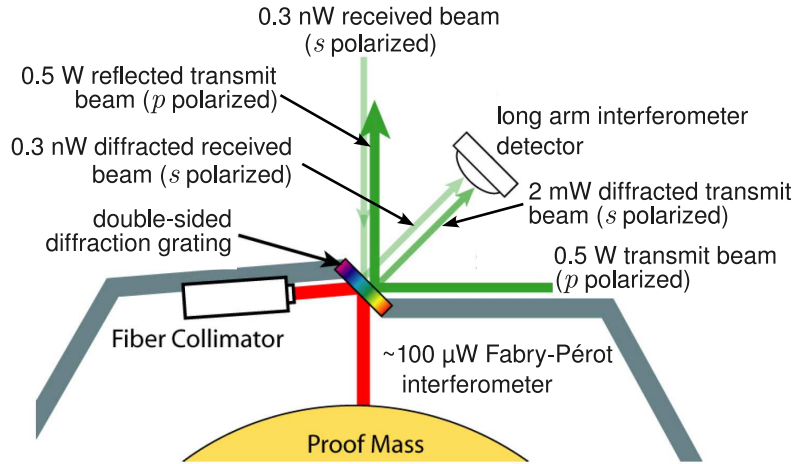}
  \caption{Primary IMS configuration using a diffraction grating
    as the reference surface and beam splitter. The readout for the short-arm
    interferometer is not shown.
    \label{fig:grating}}
  \end{center}
\end{figure}

Two interferometer configurations have been studied.
The primary configuration utilizes a double-sided diffraction grating
to act as the main reference surface and as a beam splitter as shown in
Fig. \ref{fig:grating}.
For both left and right arms, the 1/2 W beam is $s$-polarized so that
most of its power reflects off of the polarization selective,
high efficiency grating and is sent to the telescope
(dark green beam in Fig. \ref{fig:grating}).
A small portion of this beam is diffracted at 45 deg due to the
imperfect grating efficiency and strikes a photodetector.
The incoming light from the remote spacecraft is $p$-polarized
and therefore almost all ($\sim$~0.3~nW) of it is diffracted off at 45 deg
and interferes with the local beam at the detector.
The short-arm interferometer is fed from a tap-off from
the main laser using fiber,
delivering $\sim$ 100 $\mu$W of power to the underside of the grating.
This side of the grating is Littrow mounted and focusing in order to form a
Fabry-P\'erot cavity between the TM and the grating
(red beam in Fig. \ref{fig:grating}).
A finesse-enhanced Fabry-P\'erot cavity allows a direct measurement
of the distance between the grating and the TM,
without the use of a reference arm
of a Michelson interferometer.
The Pound Drever-Hall (PDH) technique \cite{PDH1983}
allows the measurement to be made at frequencies where the laser
is quantum-noise-limited.
A short-arm interferometer using a Littrow mounted grating has been
demonstrated in the lab at Stanford \cite{allen2009thesis}.

The primary interferometer configuration using a double-sided grating
has several advantages:
(a) it provides a single, well defined
reference surface that separates the long and short
arm interferometers,
(b) it can be made from a thin, low CTE material that reduces thermally
induced path-length errors relative to transmissive optics where much
higher $dn/dT$ effect are important, and
(c) it greatly reduces the number of optical components needed and decreases
the size of the optics bench.

The back-up configuration is more LISA-like,
utilizing a larger bench with bonded optical components
(see Fig. \ref{fig:IMSlayout}).
A low power portion of the 1/2 W beam for each arm
is picked off and used in both the long and
short-arm interferometers.
The short-arm interferometer is a Michelson interferometer,
while the long-arm interferometer simply interferes
the local and received beams.
A lens or mirror is used to focus the beam at the center of the TM so
that most of the light is reflected back to the interferometer.

On the frequency stabilization section of the bench, the low power
laser pick-off is fed through an
EOM which adds rf tunable sidebands
to the laser frequency.
One of the sidebands is then locked to 
a nominal 10 cm optical cavity, with a Free Spectral Range (FSR) of 1.5 GHz.
Locking is performed using PDH via
$\sim$ 10 MHz sidebands added to the main sidebands.
The main rf sidebands are tuned with a maximum range of FSR/2 in order to
shift the carrier frequency by the same amount.
Sideband locking for frequency tunable stabilized lasers has been
considered for LISA and demonstrated in a laboratory envinronment
\cite{livas2009}.
This technique is optically more efficient than using an Acousto-Optic Modulator
(AOM).
Initial tuning is performed during the initialization phase of the mission
in order to set the laser frequency on each spacecraft to the optimal offset.
The offset can then be held constant or tuned continuously to accommodate
any changes in the cavities or Doppler rates
to maintain optimal detection frequency and receiver performance.
Tuning the laser frequency offset to follow the mean Doppler frequency
of the two remote spacecraft reduces the heterodyne frequency measured
by the phase meter from $\lesssim$~150~MHz to $\lesssim$~50~MHz.

% \subsection{Phasemeter}
%
% [1 paragraph - Stanford (Conklin)]
% 
\emph{Phasemeter}:
The phasemeter measures the phase of the long-arm interferometer
beat note to $\mathrm{1 \ \mu cycle \: Hz^{-1/2}}$ over the band 1 mHz to 1 Hz.
The phasemeter uses a high-speed analog to digital converter
followed by a digital phase locked loop,
as in the LISA design \cite{shaddock2006}.
The phasemeter has to accommodate a heterodyne frequency
of $\lesssim$~50~MHz, assuming continuously tuned laser frequency offsets.
Improvements in the LAGRANGE constellation orbit may reduce this
range further.

\emph{Point-Ahead Angle Mechanism}:
A small actuator located on the optics bench
(not shown on Fig. \ref{fig:IMSlayout})
angles the output beam
with respect to the input beam $\sim$~7~$\mu$rad to accommodate the
distance that the remote spacecraft has traveled in 2.2 light sec.
path length variations must be less than 1 pm and beam jitter less than
20 nrad.
The Point-Ahead Angle Mechanism (PAAM) developed for LISA meets LAGRANGE
requirements \cite{PAAM2010}.

\subsection{Telescope}

% [0.5 design - Lockheed/Stanford (Eric Smith, David Schaechter, Dan DeBra,
%   John Lipa)]
%
The main design challenges for the LAGRANGE telescopes are:
a 5 deg field of regard (FOR) to accommodate constellation geometrical
changes,
minimized entrance aperture size to satisfy radiometric requirements,
and 5 pm pathlength stability.
Secondary requirements are $\sim$ 1 mm internal beam size for
compatibility with the metrology system,
implying a magnification of 200$\times$, and minimized stray light. 

% Telescope requirements are listed in Table \ref{tab:tel}.
%
% \begin{table}%[H] add [H] placement to break table across pages
%   \caption{\label{tab:tel} Telescope design requirements.}
%   \begin{ruledtabular}
%     \begin{tabular}{l l}
%       Design & Value \\
%       \hline \\[-2ex]
%       Field of Regard & 0 $\times$ 5 deg \\
%       Entrance aperture & 20 cm \\
%       Afocal Magnification & 200:1 \\
%       Pathlength Stability (at 1 mHz) & 5 pm \\
%       Wavelength & 1064 nm \\
%     \end{tabular}
%   \end{ruledtabular}
% \end{table}

\begin{figure}
  \begin{center}
  \includegraphics[width = 8 cm]{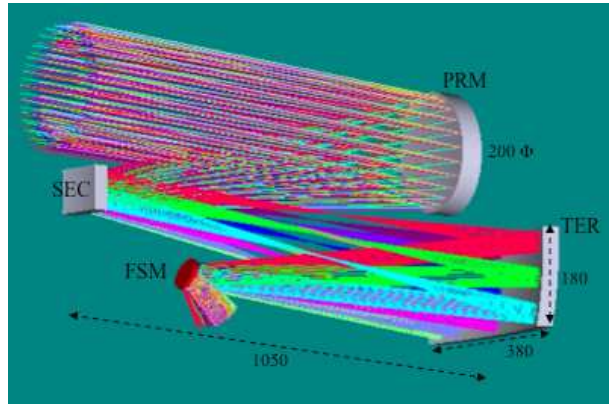}
  \caption{Stage 1 of the LAGRANGE telescope.
    \label{fig:tel}}
  \end{center}
\end{figure}

The FOR and magnification combined dictate at least a two-stage design,
and minimizing stray light, leads to an off-axis un-obscured
system that is within the capabilities of recently flown design apertures,
geometric tolerances, and wavefront stability.
Stage 1, shown in Fig. \ref{fig:tel}, is a 6:1 Three Mirror Anastigmat
(TMA) that relays the entrance pupil to a small steering mirror.
It has nearly diffraction-limited performance over the FOR, with
a steering mirror located near its exit pupil to scan over the
Field of View (FOV) (physical motion of a single steering mirror is 15 deg).
Such steering mirrors can be designed to ensure no motion of the
center of mass (CM) while slewing, and back-to-back design will also
ensure no change in CM and gravity gradient.
Optical Path Difference (OPD) induced by steering through the FOR is
$\sim$~100~pm/$\mu$rad averaged over the field.
Given a 5 deg/27.3 day (lunar orbital rate) field rate of change,
the expected OPD over 1,000 seconds is of order 3,600 pm,
which is calibrated to $10^{-3}$ or compensated at the FOR mirror.

Stage 2 gives an
additional 33$\times$ magnification and completes the 200$\times$ beam expander.
Design of this stage is relatively straightforward,
since it has a narrow FOV and the system is monochromatic.
The un-obscured TMA is a spaceflight-proven optical design from
(QuickBird \cite{quickbird1999}, MTI \cite{mti1999})
and has been used successfully by
Lockheed Martin for lasercom applications.
It may be possible to simplify the beam steering design
by taking advantage of advances in precision laser scanners
based on acousto-optic or electro-optic deflectors
that can achieve precision beam steering without moving parts
\cite{valentine2008, kremer2008}.

A low CTE composite metering structure, such as an M55J/954-3,
combined with mK temperature control achieved using heaters and
temperature sensors, can be used to control OPD
changes due to thermal effects to the pm level.

The use of the off-axis TMA for space-based imaging has significant heritage,
and is considered TRL 9.
Although mK temperature control, the use of very low CTE composite telescope
structures and pm pathlength monitoring have flight qualification test
heritage, the combination of the three to produce a pm-stable system is
new technology, and is TRL 4.

Each spacecraft has two identical telescopes.
Both are actuated so that the fixed high gain
communications antenna remains nominally pointed at the Earth,
and to provide redundancy in case of failure.

During brief periods, twice per year, the line-of-sight to the remote
spaceraft will come within 5 deg of the sun.
Existing flight-qualified narrow band filters are used to block
sunlight from entering the telescope and damaging the IMS.
A thin, low CTE window (or coating) is placed either
at the front end of the telescope, which reduces overall
solar heating to the telescope,
or at the back end, allowing for a much smaller ($\sim$ 1 cm) filter,
or possibly at both locations.

\subsection{Metrology Error Budget}

% [0.5 page - Stanford (Shally, Conklin)]
%
A detailed error budget for the IMS has been compiled.
It contains four main contributions, each with several sub-entries.
The main contributions and associated errors at 3 mHz are:
shot noise (4 $\mathrm{pm \: Hz^{-1/2}}$),
optical path-length errors (5 $\mathrm{pm \: Hz^{-1/2}}$),
residual USO phase noise (3~$\mathrm{pm~Hz^{-1/2}}$),
and residual laser phase noise (3~$\mathrm{pm~Hz^{-1/2}}$).
The total error at 3 mHz is 8 $\mathrm{pm \: Hz^{-1/2}}$.
The IMS error is nearly flat at higher frequencies, and exhibits a $1/f^2$
trend below 3 mHz.

The shot noise is limited by the laser output power (0.5 W),
arm length (670,000~km) and telescope aperture (20 cm),
while the optical path length error
is dominated by the telescope design with its required 5 deg beam steering.
The residual USO phase noise calculation is based on the shot noise limit,
the phasemeter error, a heterodyne frequency of 50 MHz, 
and a fractional arm length difference of 5\%.
The residual laser phase noise estimate assumes a
USO clock Allan deviation of $\mathrm{5\times10^{-11}}$ at 4 sec
(round-trip light travel time),
an absolute arm length uncertainty of 10 m,
and a pre-stabilized laser frequency
noise of $\mathrm{30 \ Hz \: Hz^{-1/2}}$ at 3 mHz.

\section{Disturbance Reduction System}

The disturbance reduction system consists of the Modular Gravitational
Reference Sensor, which houses the test mass, drag-free and
attitude control laws and micronewton thrusters
to keep the S/C centered on the test mass, and a thermal control system.

\subsection{Modular Gravitational Reference Sensor}

Based on experience with the successful drag-free satellites
Triad I \cite{triad1974} and Gravity Probe B \cite{bencze2006},
a spherical geometry was chosen for the LAGRANGE GRS.
A spherical GRS for LISA was
proposed as early as 1998 \cite{debra1999, keiser2000}, and
has advantages that outweigh its disadvantages:
\begin{itemize}
  \item No TM forcing or torquing: neither electrostatic support nor
    capacitive sensing is required, reducing disturbances and complexity,
  \item Large TM-to-housing gap (35 mm): disturbances are reduced and
    spacecraft requirements are relaxed,
  \item A long flight heritage \cite{debra2011}: Honeywell gyroscopes,
    Triad I and GP-B,
  \item Scalability: performance can be scaled up or down by adjusting
    TM and gap size,
  \item Simplicity: no cross coupling of degrees of freedom,
  \item A simple flight-proven caging mechanism.
\end{itemize}

\begin{figure}
  \begin{center}
  \includegraphics[width = 12 cm]{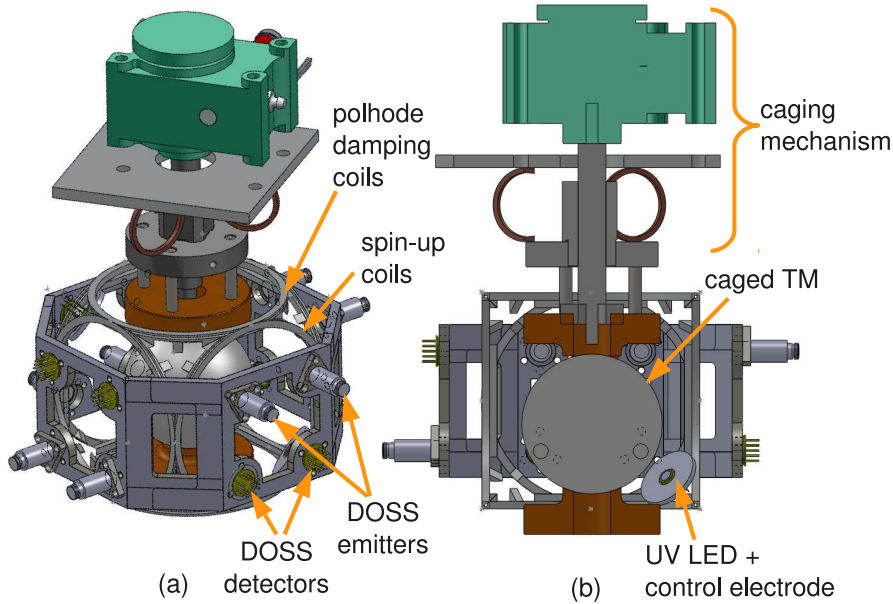}
  \caption{(a) Modular Gravitational Reference Sensor isometric view, and (b)
    cross-section with the test mass caged.
    \label{fig:MGRS}}
  \end{center}
\end{figure}

This GRS concept, now called the Modular Gravitational Reference Sensor (MGRS)
has been under development for a wide range
of applications since 2004 \cite{GRSwhitePaper2004, sun2006aip}.
The primary components of the MGRS, shown in Fig.~\ref{fig:MGRS},
include a spinning spherical TM,
a differential optical shadow sensor system for drag-free control,
a caging (launch lock) mechanism based on the flight-proven
DISCOS design,
magnetic coils for test mass spin-up to 3-10~Hz and polhode damping,
based on the Honeywell design,
and a charge control system based on the GP-B design but using
modern LEDs as UV sources.

As an alternative,
we note that the LISA Pathfinder GRS \cite{pathfinderGRS2003},
the baseline for LISA, is expected to demonstrate
$\mathrm{3\times10^{-14} \ m/s \: Hz^{-1/2}}$ above 1 mHz
during the LISA Pathfinder mission \cite{LPFstatus2011}.
Assuming a successful flight demonstration, the Pathfinder GRS
could be utilized for LAGRANGE.
A single LISA Pathfinder GRS per drag-free spacecraft has been studied
\cite{gerardi2012} and would principally require modifications to the
optical bench design.

% \subsubsection{Spinning Spherical Test Mass}
%
% [1 page - Stanford (Conklin, DeBra, Karthick)]
% 
\emph{Spinning Spherical Test Mass}:
The nominal test mass is a 2.9 kg, 70 mm diameter sphere of
70\%/30\% Au/Pt.
An alternate material is Berglide (2\%/97.5\%/0.5\% Be/Cu/Co),
a common, well studied material,
easier to fabricate with $10^{-6}$ magnetic susceptance,
but less dense by a factor $>$~2.

The TM must be round to $\lesssim$ 30 nm, similar in roundness to the
GP-B flight rotors \cite{keiser2009},
and have a mass unbalance $\lesssim$ 300 nm, 30 times larger
than that of the GP-B rotors.
Internal axi-symmetric sections of the TM are hollowed out to produce
a moment of inertia difference ratio of 10\%, while reducing the average
density by 20\% \cite{conklin2009thesis}.
The resulting polhode frequency is 0.3-1 Hz, at the high end of the LAGRANGE
science band.
The hollow sections also allow the 
mass center to be moved within the sphere through
an iterative measurement/re-shaping procedure
in order to meet the mass unbalance requirement \cite{conklin2009thesis}.
Laboratory measurement of the mass unbalance of a 50 mm spherical TM has
been demonstrated to 200 nm \cite{conklin2011}.
For a GRS housing vented to space, the pressure is $10^{-6}$ Pa, resulting
in a spin-down time $\sim$~4,000 years.

Interferometric measurement of the surface of a sphere has also
been demonstrated in the lab \cite{allen2009thesis}, and
spinning of the sphere averages geometric irregularities,
allowing for determination of the mass center.
A computationally simple and robust on-board algorithm for
determining the mass center to pm accuracy has been
developed analytically \cite{conklin2008jgcd} and
demonstrated numerically \cite{allen2008gcd}.
The TM external geometry requires no special markings
or cut-outs, and pm-level knowledge of the sphere's geometry is not
needed.

The TM spin axis is normal to the constellation plane
($\pm$ 5 deg) to achieve maximum averaging of geometric irregularities.
Systematic measurement errors due to
the axisymmetric harmonics of the sphere's geometry remain below 1 pm
as long as the out-of-plane spacecraft motion is
$\mathrm{< 3 \: \mu rad \: Hz^{-1/2}}$ and
the maximum rate is $\mathrm{< 0.6 \: \mu rad/s}$ \cite{conklin2008jgcd}.
The former requirement is bounded by the 
$\mathrm{< 10 \: nrad \: Hz^{-1/2}}$ attitude motion requirement for the IMS,
and the latter is greater than the maximum attitude rate
($\mathrm{\sim \: 0.2 \: \mu rad/s}$)
needed to maintain pointing throughout the orbit.

The TM is coated in a carbide compound (e.g. SiC or ZrC),
which provides a hard, conductive, and highly reflective surface.
SiC and ZrC have quantum efficiencies 15-30\% of that of gold,
thus supporting UV charge control.
A major advantage of these coatings over gold is their low adhesion
to other surfaces.
The coating of the TM and MGRS, which has no sensitive surfaces,
are designed such that the TM can repeatedly
touch the housing wall in a $\mu$g environment without damaging the TM
or housing or sticking.
This greatly simplifies the caging design and means that during
low thrust station keeping maneuvers (once every 6-12 months), the TM 
need not be re-caged, but only spun down to a low level.

% \subsubsection{Shadow Sensor for Drag-free Control}
%
% [2 paragraphs - Stanford (Andreas)]
%
\emph{Shadow Sensor for Drag-free Control}:
The Differential Optical Shadow Sensor (DOSS) requires two pairs of
parallel beams for a three-dimensional position measurement
\cite{zoellner2011}.
Four pairs are planned for redundancy.
Superlumniscent LEDs with a wavelength of 1550 nm are
used for the light source.
The target sensitivity is 1~$\mathrm{nm \: Hz^{-1/2}}$
over the drag-free control bandwidth.
The low frequency noise floor is
improved with lock-in amplification and modulation of the beam power.

% \subsubsection{Caging}
% 
% [2 paragraphs - Stanford (Eric)]
%
\emph{Caging}:
The function of the caging system is to secure the test mass during launch
and ascent when the satellite can be expected to experience high
static accelerations (up to 6 g \cite{falcon9}),
random vibration up to 14.1 g rms \cite{GSFC-STD-7000},
and shock up to 3000 g at payload separation \cite{falcon9}.

The caging mechanism for the MGRS is based on the flight proven design for
DISCOS, which applied a load of 43 g to the TM \cite{hacker1976}. 
The caging system will passively maintain a high compression pre-load
until deployment in orbit
when the TM will be released with low residual velocity.
The TM and housing are designed such that they will contact
several times with $\sim$ mm/s velocities, with no damage, until sufficient
kinetic energy is absorbed and the drag-free control
system is able to capture the TM.
% It is desirable that the caging system also provide a means to exercise
% three axes of the optical position sensor by moving the test mass,
% and have the ability to re-cage the test-mass.

The test mass is clamped between two holding tubes,
one of which is driven by a trapezoidal jack screw
(see Fig. \ref{fig:MGRS}).
% The contact area for each holding tube consists of a spherical 
% annulus broken into two sections.
% The holding tubes are oriented so that the contact patches on
% four meridians at equal intervals around the sphere.
% This configuration hedges against the eventuality that the holding
% tubes will not be precisely coaxial.
The surface finish for the holding tubes and the TM will be
of dissimilar materials to avoid residual adhesion.
The jack screw is self-locking to passively maintain compression force,
and is driven by a piezoelectric motor which is also self-locking.
% Commercially available positioners can produce a small amount of torque
% (for example, the PI M-660 can produce 0.3 Nm# and will operate in a vacuum).
% A gear box is used to couple the piezomotor to the jack screw.
% For reference, with a gear ratio of 20:1 and a no-load torque of 0.11 Nm,
% the Nook Industries screw jack 16 mm in diameter with a lead of
% 4 mm could develop more than 500 N of force#.
% The jack screw will use a lubricant which is compatible
% with ultra-high vacuum.
% The lead screw will be engaged before satellite integration.
%Precisely-machined
BeCu ring springs oppose the compression force
of the jack screw, such that
at full load, the opposing surfaces that retain the springs will
engage a limit switch, and stop the jack screw.

% \subsubsection{Spin-up Mechanism}
%
% [2 paragraphs - Stanford (Andreas)]
%
\emph{Spin-up Mechanism}:
The spin-up of the TM is performed with a rotating magnetic
field similar to the ones used in Honeywell
gyroscopes \cite{lange1964thesis}.
Four magnetic coils separated by 90 deg in the constellation plane
are excited with ac currents to create
the rotating magnetic field and will perform spin-up within a few hours.
Two additional magnetic coils aligned normal to the constellation plane
create a dc field for
polhode damping and spin-axis alignment (see Fig. \ref{fig:MGRS}).
The TM can also be spun-down by reversing the phase of the
ac currents.

% \subsubsection{Charge Control}
%
% [0.5 pages - Stanford (Sasha, Karthick)]
%
\emph{Charge Control}:
Charge management by UV photoemission using the 254 nm line of an rf
mercury source was successfully demonstrated by the GP-B mission in 2004-2005.
Newer  technology allows the use of commercially available LEDs
operating in the 240-255 nm range \cite{sunpatent} as the UV source.
These devices are fast switchable ($f~>$~100~MHz),
allowing pulse timing to be synchronized to a control electrode.
With a 10 mA driving current, these LEDs are capable of generating
10 $\mu$W at 252 nm \cite{balakrishnan2011}.
Electrons are generated through photoemission
from the TM and control electrode.
The direction of charge transfer is selected by setting the phase between
the UV-LED and control electrode \cite{sunpatent}.
% When the LED is in phase with the electrode, generated electrons are
% pulled away from the test mass and towards the housing
% resulting in positive charging.
% Similarly, when the LED is out of phase with the control electrode,
% generated electrons are pushed towards the TM,
% resulting in negative charging.
Measurement of the TM potential can be performed in several ways,
including the force modulation used in GP-B and
contactless dc measurement of the electric field.
Passive charge management, relying on a virtual ``wire''
generated by photoemission and without bias is also practical for the
proposed low ($\sim$~5~pF) capacitance MGRS.
The power and mass per GRS are estimated at 2-3~W and 200-300~g, respectively.
A number of UV-LED models have successfully completed environmental
testing \cite{balakrishnan2011, sun2009}.

% The UV source for charge management \cite{buchman1995} will use a UV-LED
% available in the 240-255 nm range \cite{sunpatent}.
% These devices are fast switchable ($>$ 100 MHz) and will  be synchronized
% to the local electrodes frequency.
% The phase between the UV-LED supply and the electrode normal excitation
% will determine the polarity of charge control.
% The power and mass per GRS are estimated at 2-3 W and 0.2-0.3 kg.
% Fiber coupled UV-LED's are available making the new interface with the
% LPF GRS  compatible with the old ones. A number of models of UV LED's
% have been successfully environmentally qualified on the ground \cite{sun2009},
% and a small satellite demonstrating the entire charge management
% system in space will be launched in 2012-2013.

\subsection{Drag-free and Attitude Control}

% [1 paragraph - Stanford (Conklin)]
%
Three-axis drag-free translation control keeps the spacecraft
centered on the TM to within $\mathrm{2 \ nm \: Hz^{-1/2}}$
in the measurement band.
Each axis is controlled independently (no cross-coupling).
The drag-free position accuracy is limited by the DOSS noise
($\mathrm{1 \ nm \: Hz^{-1/2}}$) and
the dominant spacecraft disturbances, which are solar radiation pressure
at 1 AU ($\mathrm{\lesssim 10^{-10} \ m/s^2 \: Hz^{-1/2}}$
\cite{schumaker2003cqg}) and
micronewton thruster noise ($\mathrm{3 \times 10^{-10} \ m/s^2 \: Hz^{-1/2}}$).

The 3-axis attitude control is completely independent from the drag-free
control and aligns the two
telescopes to the two remote spacecraft using wavefront sensing as in LISA.
% The required accuracy is $\mathrm{10 \ nrad \: Hz^{-1/2}}$
% in the measurement band.
The remaining degree of freedom is accommodated by telescope beam steering
using the steering mirror mentioned above.

In addition to the 3 drag-free and 3 attitude degrees-of-freedom, there
is the telescope breathing angle control yielding a total of 7
controlled degrees-of-freedom for each LAGRANGE spacecraft.
This is significantly less than LISA which must control
a total of 19 degrees-of-freedom per spacecraft.

\subsection{Micronewton Thrusters}

% [2-3 paragraphs - SRI]
% 
% \emph{Micronewton Thrusters}:
Drag-free translational control and spacecraft pointing are both
actuated by a micronewton electric propulsion system.
The requirements for precision and noise are equivalent to those for LISA:
0.1 $\mathrm{\mu N \, Hz^{-1/2}}$ thrust noise from 1 mHz to 1 Hz and
0.1 $\mathrm{\mu N}$ thrust precision.
Busek Colloid Micro-Newton Thrusters (CMNT) meet both of these requirements.
Further development is required to meet the 5 year lifetime goal
\cite{busek2008a, busek2008b}.

In addition to the Busek CMNTs,
two Field Emission Electric Propulsion (FEEP) systems
have been investigated for
LISA and LISA Pathfinder, a caesium slit FEEP \cite{italianFEEP2009}
and an indium needle FEEP \cite{austrianFEEP2011},
which was also considered for the GOCE mission \cite{austrianFEEP2004}.
However, no thruster has
thus far demonstrated all requirements for noise,
precision, dynamic range in thrust and lifetime \cite{LPFstatus2011}.
A significant effort in micronewton propulsion
technology development and testing is needed.

An attractive alternate thruster baselined for LAGRANGE is a
scalable ion propulsion concept based on micro-fabricated arrays of liquid
metal ion sources, currently under development at
SRI International \cite{aguero2001, aguero2003}.
Thrust is generated by the acceleration and control of independently created
ions and electrons, each generated using arrays of micro-fabricated
emission sites.
The use of independent extraction and acceleration electrodes enables
very high mass efficiency (high specific impulse)
and wide dynamic range of thrust.
This control approach allows smooth variation of thrust over the
full operating range.

\begin{figure}
  \begin{center}
  \includegraphics[width = 6 cm]{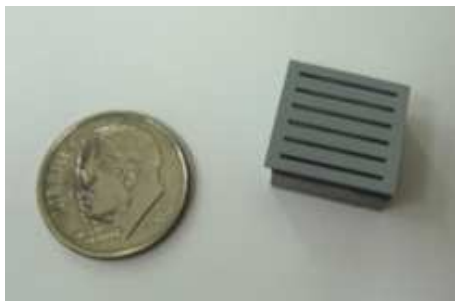}
  \caption{SRI fabricated
    linear array version of the liquid metal ion source with 1 $\mathrm{cm^2}$
    active area.
    Prototypes were operated $>$ 30 hours, with multiple
    start/stop, and even atmospheric exposure between runs to test device
    robustness and physics.
    \label{fig:SRIion}}
  \end{center}
\end{figure}

Prototype ion sources have been operated in the 1-5~W
range, and ion source operation has been validated from 2,000 s to
10,000 s specific impulse.
A prototype ion source is shown in Fig. \ref{fig:SRIion}.
Because of the lower operating voltages enabled
by microfabrication of ion emission sites, packaging,
including control electronics and power conversion, are
expected to occupy $\mathrm{<~10~cm^2}$.
Arrays of up to 160 emitters have been tested, with prototypes able to
handle up to 480 emitters;
each emitter in such an array is capable of approximately 1-10~nN
of thrust and can be pulsed to produce pN-sec
impulse bits if each emission site is independently controlled.

The micro-fabricated scale of the elements 
results in rapid neutralization of the particle streams to permit
high currents from such a small area device.
Hence, this developing technology can be scaled to arrays of arbitrary
size to provide nanonewtons to newtons of thrust while meeting all of the
observatory propulsion requirements.
For example, this allows these thrusters to be used for both drag-free
operations and for the $\sim$ 1 m/s station keeping maneuvers needed
once every 6 to 12 months.

\subsection{Thermal Control System}

% \emph{Thermal Control System}:
%
Spacecraft heating around the outside surface varies at the orbital
period of 27.3 days.
The payload is kept nominally at 300 K, while the exterior sun-facing solar
arrays heat to roughly 350 K.
The solar arrays are thermally isolated to keep exterior spacecraft components
stable to 1~K, and additional thermal control on the telescopes keeps
them stable to 1 mK.

The 27.3 day thermal cycle is $10^3$
below the minimum science frequency of 1 mHz.
This greatly
reduces the thermal impact on the science instrument,
which requires 10~$\mathrm{\mu K \: Hz^{-1/2}}$ 
temperature stability in the science band.
To isolate the MGRS and optics bench from the $\sim$~1~K
spacecraft temperature
variations, a nominal thermal enclosure consisting
of a 2-4 alternating layers of
highly conductive shields and vacuum spacing is employed.
Radiative heat transfer can be further reduced by coating with
low emissivity materials.
Shiny gold coating reduces emissivity to $\sim$ 0.02 \cite{jin2006},
3.5 times less than a rough surface.
A thermal control system with $<$~$\mu$K stability has been designed
with COMSOL for the low earth orbiting STAR spacecraft,
validating the concept\cite{higuchi2009, alfauwaz2011}.

\subsection{Acceleration Noise Budget}

\begin{figure}
  \begin{center}
  \includegraphics[width = 10 cm]{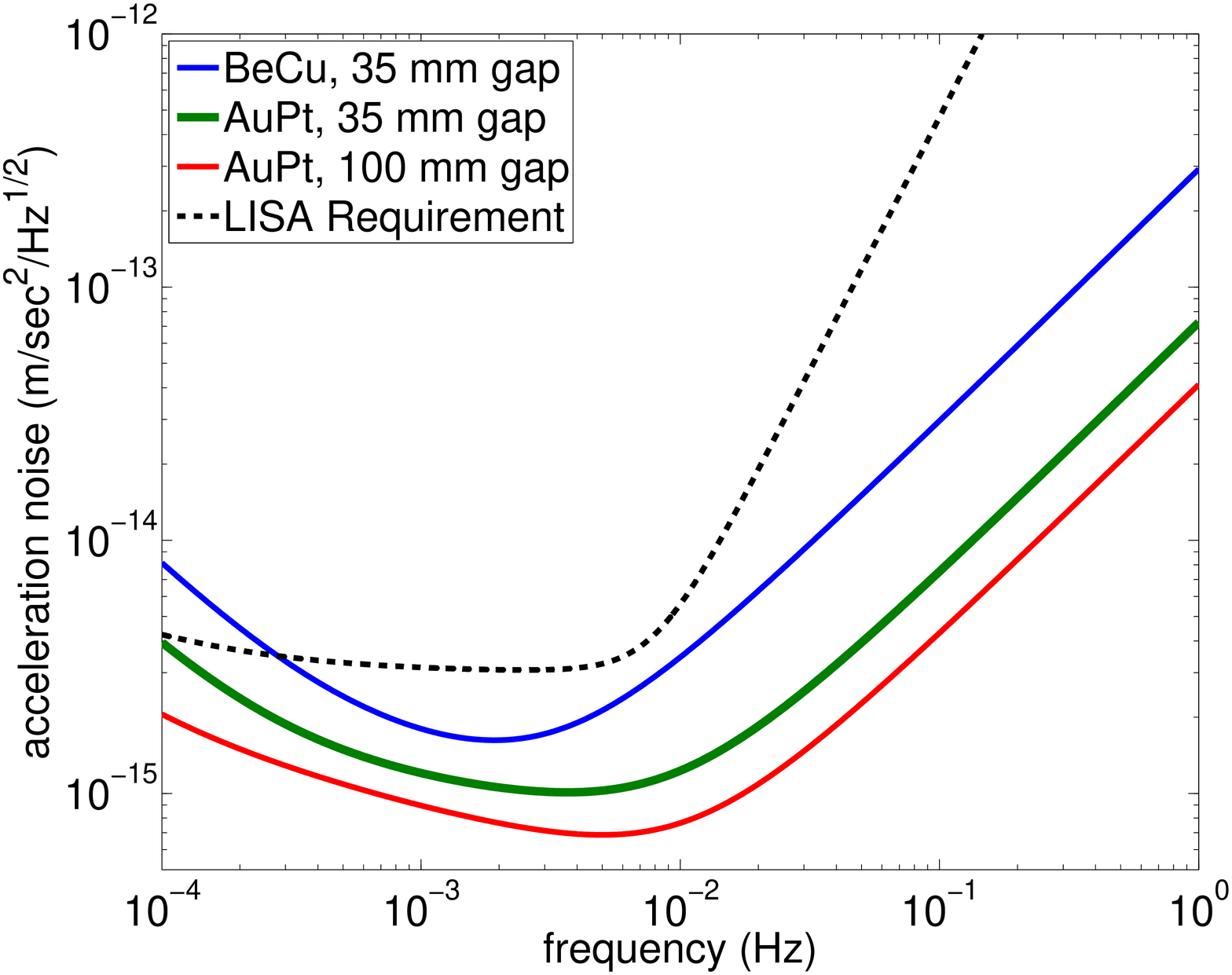}
  \caption{Estimated MGRS acceleration noise performance for a 35 mm diameter
    AuPt test mass with a 35 mm gap (green), with the gap increased to 100 mm
    (red), and with a Berglide TM (blue).
    The LISA requirement is shown for comparison. 
    \label{fig:acclNoise}}
  \end{center}
\end{figure}

% [0.5 page - Stanford (Conklin)]
% 
A detailed acceleration noise budget has been compiled for the MGRS.
The budget contains 30 terms: 6 S/C-to-TM stiffness, 
8 magnetic, 6 thermal, 4 electric, 4 Brownian, 1 cosmic ray,
and 1 laser noise term.
Calculation of each term in the acceleration noise budget
follows the methodology
used for LISA \cite{schumaker2003cqg, gerardi2012}.
The resulting composite acceleration amplitude spectral density
is shown in Fig. \ref{fig:acclNoise} for three possible configurations.
The baseline design consists of an AuPt sphere, 70 mm in diameter with a
35 mm gap between TM and housing
(green curve in Fig. \ref{fig:acclNoise}).
Also shown are the results the same geometry, but with
a Berglide sphere (blue) and for
an AuPt sphere with a 100 mm gap (red),
demonstrating the performance scalability of the MGRS.

The dominant acceleration noise contributions below 0.5 mHz are
due to TM charge and stray (patch) voltage interactions.
Between 0.5 mHz and 2 mHz the composite
noise is dominated by TM-to-spacecraft gravitational
interactions, and above 2 mHz by S/C magnetic field fluctuations.

The TM-to-spacecraft gap size, $d$, is the most important design
parameter with respect to acceleration noise performance.
Magnetic, Electric, and the largest of the Brownian disturbances
\cite{weber2009}
are proportional to $1/(\rho d)$ ($\rho$~=~TM density).
The MGRS gap size (35 mm) and TM density ($\mathrm{2\times10^4 \ kg/m^2}$)
are 1,000 and 10 times larger than that of
GP-B respectively.
As a cross-check, we scale the acceleration noise performance of GP-B
($4 \times 10^{-11} \ \mathrm{m/s^2 \: Hz^{-1/2}}$
from 0.01 to 10 mHz \cite{bencze2006}) by these ratios and obtain
for the MGRS, an acceleration noise of
$4 \times 10^{-15} \ \mathrm{m/s^2 \: Hz^{-1/2}}$.

% \begin{figure}
%   \begin{center}
%   \includegraphics[width = 16 cm]{acclNoiseDetail_R3.eps}
%   \caption{Term-byterm acceleration noise error budget for the baseline design
%     (70 mm diamter AuPt sphere with 35 mm gap). 
%     \label{fig:acclNoise}}
%   \end{center}
% \end{figure}

When constructing the acceleration noise budget,
environmental requirements below 1 mHz 
for spacecraft-to-TM stiffness,
spacecraft temperature fluctuations,
stray voltages, TM charge,
and spacecraft magnetic environment were relaxed relative to
LISA \cite{schumaker2003cqg, gerardi2012}, some by
an order of magnitude in order to simplify the
spacecraft and MGRS design and reduce cost.
This can be seen in Fig. \ref{fig:acclNoise} for example,
where below 0.5 mHz electrostatic disturbances,
proportional to 1/$d$, dominate.
The LAGRANGE gap size (35~mm) is roughly 10$\times$ larger than that of LISA,
and yet the LAGRANGE baseline design (green curve in Fig. \ref{fig:acclNoise})
is roughly equivalent to the LISA requirement (black, dashed curve).

\section{Spacecraft and Mission Design}

\subsection{Spacecraft Design}

%[0.5 pages - NASA ARC/Lockheed (Hugo Sanchez lead, support from LM)]
%
The LAGRANGE spacecraft is based on an existing Lockheed configuration
that has flown successfully many times.
Indeed, Lockheed's vast experience, with more than 950 spacecraft flown,
adds confidence to the overall LAGRANGE mission success.
The spacecraft is a dodecagon structure $\sim$3 m in diameter and $\sim$0.7 m
tall, consisting
of a compact equipment section
with an inner diameter center bay which accommodates the telescopes and
payload.
A fixed high gain antenna is mounted between the two telescopes.
To minimize the telescope and payload deformations, the spacecraft material
will be thermally controlled in order to maintain a low
thermal gradient and a high level of thermal stability.
The spacecraft equipment is mounted in the outer bays and oriented
to minimize pointing error.
The solar arrays consist
of fixed panels mounted on the outer sides of the spacecraft structure.
Radiators are located on the top and bottom of the spacecraft
for thermal control.   

The total mass of each LAGRANGE vehicle is $<$ 470 kg, including payload,
requiring $<$ 500 W of power while transmitting data to the ground.
% yielding a 33\% launch mass margin with the propulsion module.
The mass estimate for each component has been evaluated and assigned a
maturity rating, and a contingency value assigned from Lockheed's standard
weight/power growth allocation and depletion schedule based on history
and experience from actual measurements of flight hardware.
% The hardware contingency factor is reduced incrementally as the maturity
% improves.
% The subsystem, spacecraft, payload, and satellite contingency totals are
% the sums of the contingency values used for the constituent components. 

The spacecraft are designed for a minimum of five years of operation,
easily complying with the baseline LAGRANGE mission duration
requirement.
% The spacecraft incorporates a robust design capitalizing on advances
% in heritage electronics design and increased reliability
% and augmented with a conservative test program.

\subsection{Communications}

% [1 paragraph - Stanford (Ahmad) with crosscheck from Lockheed]
% 
Communications with each spacecraft is performed with
a direct link to ground stations.
LAGRANGE will use the standard NASA ground network, consisting of 10-11 m
antennas located all over the Earth.
The down-link rate is 1 Mbps and the up-link rate 1 kbps,
with a 9.6 ratio of received energy per-bit to noise-density.
The transmitter power is 10 watts and the half power beam width is $>$8~deg.
An advantage of this design is that it allows near
continuous communications with all three spacecraft
during the initialization phase and other critical phases of the mission.

\subsection{Mass and Power Budgets}

% [1 paragraph + 1 table - NASA ARC (John Hansen lead)]
%
Table \ref{tab:budget} shows the mass and power requirements for
each of the three spacecraft, as well as the total wet launch mass,
including the propulsion module and launch adapter.
The propulsion module carries 230 kg of bi-propellant with
an Isp of 320 s.
This provides a total $\Delta v$ of 600 m/s to the three spacecraft
stack plus an additional 30\% margin.

\begin{table}%[H] add [H] placement to break table across pages
  \caption{\label{tab:budget} Spacecraft and Payload mass and power budget.}
  \begin{tabular}{l c c}
    \hline
    & Mass* (kg) & Power** (W) \\
    \hline
    \textbf{Spacecraft ($\times$ 3)} & & \\
    Payload & 170 & 175 \\
    Spacecraft & 300 & 325 \\
    Total spacecraft + payload & 470 & 500 \\
    \textbf{Propulsion module ($\times$ 1)} & & \\
    dry propulsion module & 330 & \\
    propellant & 230 & \\
    \textbf{Launch adapter ($\times$ 1)} & 100 & \\
    \textbf{TOTAL} & \textbf{$<$ 2,070} & \\
    \hline
  \end{tabular}
  \flushleft * including 30\% margin for payload and propulsion module,
    14\% margin for $>$ TRL 6 Lockheed spacecraft\\
    ** including 30\% margin for payload, 8\% margin for 
    spacecraft, while transmitting\\
\end{table}

The baseline launch vehicle is the Falcon 9 Block 2 with a
4.6 m diameter $\times$ 6.6 m tall fairing.
% The cost through the NASA Launch Services Program is roughly \$118 M,
% including NASA overhead.
The maximum launch mass for characteristic energy,
$\mathrm{C3 = 0 \ kg^2/s^2}$, is 2,500 kg.
The total wet launch mass, $<$~2,070 kg, which includes 30\% margin,
makes up only 80\% of the total capacity of the Falcon 9.
The mass, stack dimensions and C3 are all consistent with
the more energetic Atlas V 401 and possibly smaller, less expensive
launch vehicles.

\begin{figure}
  \begin{center}
  \includegraphics[width = 8 cm]{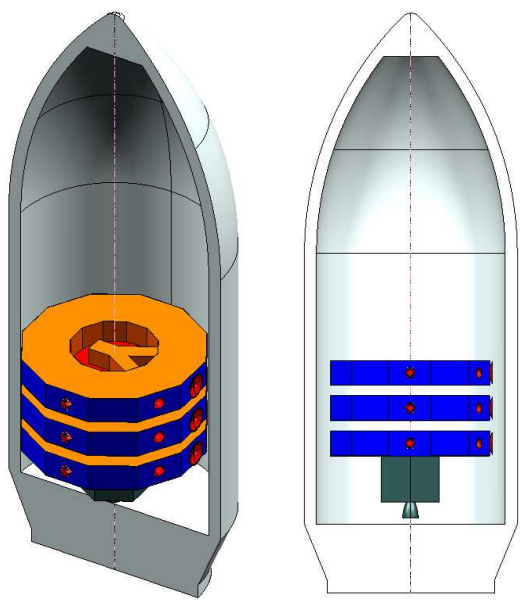}
  \caption{Conceptual view of the
     three LAGRANGE spacecraft plus propulsion module inside
    the Falcon 9 fairing. 
    \label{fig:fairing}}
  \end{center}
\end{figure}

\subsection{Mission Design}

% [0.5 pages - NASA ARC (Anthony, Hugo)]
%
%
% \begin{figure}
%   \begin{center}
%   includegraphics[width = 8 cm]{earthMoonLAGRANGE.eps}
%   \caption{Launch and orbit insertion trajectory}
%     \label{fig:phaseOrb}
%   \end{center}
% \end{figure}
%
The LAGRANGE mission is divided into four phases:
(a) launch plus $\sim$ 6 month cruise to the science orbits
(after which the propulsion module is ejected),
(b) Initial Orbit Checkout (IOC),
which includes starting drag-free operations
and acquisition of the signal from the remote spacecraft,
(c) 5-year science observations,
and (d) de-commissioning at the end of the mission.

The three spacecraft are stacked together with a single
propulsion module inside the launch vehicle fairing
as shown in Fig. \ref{fig:fairing}.
After separation from the launch vehicle upper stage, the propulsion
module brings all three spacecraft into a phasing orbit, which lies
in the plane of the Moon's orbit with apogee at the lunar apogee
($\sim$~384,000~km) and eccentricity to achieve a 33 day orbital
period.
% (see Fig. \ref{fig:phaseOrb}).
This orbit is designed so that at apogee, the propulsion module
and 3 spacecraft return to the Moon's orbit with a 60 deg phase
shift every 33 days.
Every 66 days one of the three spacecraft is delivered into its Lagrange
point.

Two types of low thrust injections have been identified that can achieve this
phasing orbit: 1) The baseline is a
direct launch to the Weak Stability Boundary (WSB) \cite{WBS1987},
followed by a small $\Delta v$ to return to the Earth-Moon system,
and then a lunar swing-by; 2) the alternate is a
direct launch into Trans-lunar Injection (TLI) \cite{clarke1963} followed
by a lunar swing-by coupled with a larger $\Delta v$ from the propulsion module.
The WSB injection lasts 9 months, requiring a C3 of 0 $\mathrm{kg^2/s^2}$
and a 600 m/s total $\Delta v$ from the propulsion module,
while the TLI injection takes only 6-7 months and
requires a C3 of $- 1.7 \ \mathrm{kg^2/s^2}$
and a 800 m/s total $\Delta v$ from the propulsion module.

After reaching the science orbit, the mission lifetime is planned to be
5 years, and is limited only by the lifetime of the science instrument
and the micronewton thrusters.
The Earth-Moon L3, L4, L5 orbit can be maintained indefinitely with
$\sim$ 1 m/s $\Delta v$ every 6-12 months for station keeping. 
During these maneuvers, the test mass must be spun-down, but not re-caged
for accelerations $\mathrm{\sim 10^{-5} \ m/s^2}$.

LAGRANGE data analysis process would proceed as in LISA,
although access to the science data would occur closer to real-time
compared with LISA due LAGRANGE's orbit and reduced data latency.
Phase measurements from each spacecraft are combined using TDI
\cite{armstrong1999} and stored on public networks for analysis
by remote science investigators.
Existing LISA data analysis methodologies
\cite{mockLISAda2008} would directly apply to the LAGRANGE data.
The main difference would be the change of antenna pattern due
to the different orbit.

\subsection{Order-of-Magnitude Cost Estimate}

% [2 Paragraphs + 1 Table - NASA ARC (John Hansen, Dohy?)]
%
LAGRANGE cost is estimated in the medium range of \$600M to \$1B.
A detailed and conservative joint ARC and Lockheed Martin cost analysis
puts the mission ROM
cost at \$950M FY12, including 30\% reserves.
LM has orbited over 950 S/C, supporting many relevant programs that have
segments, subsystems or components similar to LAGRANGE and were used to
get actual cost data.
The nonrecurring costs for the development of 3 identical
spacecraft with hardware elements at TRL 6 or greater
have been accounted for.
The payload ROM cost was developed using a combination of bottoms-up and
analogies based on major components.
ROM cost for the remaining mission elements used a
wrap factor applied to the base hardware costs (spacecraft and payload),
which accounts for systems engineering and testing by an
industrial contractor and government costs.
The base hardware costs were modeled on similar size and complexity missions
and derived from the NASA ARC cost database of these missions.
Simplistically using the function describing the historical mission cost
data we obtain estimates in the \$640M to \$895M FY 11,
depending on assumptions about the cost of 3 identical systems.

\section{Technology Readiness}

% [0.5 pages - Stanford/Lockheed team]
%
Table \ref{tab:TRL} shows the flight heritage and TRL for each of the main
LAGRANGE sub-systems.
Except for a few sub-components,
all sub-systems are at TRL 5 or higher.

\begin{table}%[H] add [H] placement to break table across pages
  \caption{\label{tab:TRL} Payload \& spacecraft heritage and TRL.}
  \begin{tabular}{l c c}
    \hline
    Component & Heritage & TRL \\
    \hline
    Spacecraft & Lockheed & $>$6 \\
    Laser system & LISA Pathfinder \cite{LPFstatus2011} & 6 \\
    Charge control & UV-LED Sat  \cite{balakrishnan2011} & 6 \\
    GRS & DISCOS \cite{triad1974}, GP-B \cite{everitt2011},
      LISA \cite{LPFstatus2011} & 5 \\
    Laser freq. system & LISA \cite{LPFstatus2011}, STAR & 5 \\
    Phasemeter & GPS, LISA \cite{shaddock2006} & 5 \\
    Caging mechanism & DISCOS  \cite{hacker1976} & 5 \\
    $\mu$N thrusters & SRI  \cite{aguero2001}
      (CMNT\cite{busek2008b}), In FEEP\cite{austrianFEEP2004}) & 4(6) \\
    Telescope & QuickBird \cite{quickbird1999},
      MTI \cite{mti1999} & 4 \\
    PAAM & LISA \cite{PAAM2010} & 4 \\
    Shadow sensor & SALKS small sat. \cite{zoellner2011} & 4 \\
    \hline
  \end{tabular}
\end{table}

\section*{Acknowledgments}

We thank
Leo Hollberg (Stanford) for sharing
knowledge and experience with optical cavities
and laser interferometers,
Ke-Xun Sun (University of Nevada, Las Vegas)
for several technical contributions,
Stefano Vitale (University of Trento) for insights into the LISA
and LISA Pathfinder GRS,
Ron Hellings (Montana State University) for engineering and science
implications of geocentric gravity-wave observatories and telescope filters,
Cesar Ocampo (University of Texas, Austin)
for initial information regarding Earth-Moon Lagrange points,
Vlad Hruby (Busek Co, Inc) for input on FEEPs and their spacecraft integration,
and the entire LISA team for their contributions to this field
over the past few decades.

\section*{References}

\bibliography{LISA_RFI_2011}

\end{document}